\documentclass{ws-ijmpa}
\usepackage[super,compress]{cite}
\usepackage{graphicx}
\usepackage{graphicx}
\usepackage{dcolumn}
\usepackage{bm}
\usepackage{epstopdf}
\newcommand{\hs}{\hspace*{0.5cm}}

\newcommand{\be}{\begin{equation}}
\newcommand{\ee}{\end{equation}}
\newcommand{\bea}{\begin{eqnarray}}
\newcommand{\eea}{\end{eqnarray}}
\newcommand{\ben}{\begin{enumerate}}
\newcommand{\een}{\end{enumerate}}
\newcommand{\bde}{\begin{widetext}}
\newcommand{\ede}{\end{widetext}}
\newcommand{\nn}{\nonumber}
\newcommand{\crn}{\nonumber \\}

\newcommand{\al}{\alpha}

\newcommand{\om}{\omega}

\newcommand{\fr}{\frac}
\newcommand{\bc}{\begin{center}}
\newcommand{\ec}{\end{center}}

\newcommand{\De}{\Delta}

\newcommand{\bit}{\begin{itemize}}
\newcommand{\eit}{\end{itemize}}
\begin{document}

\catchline{}{}{}{}{}

\title{Lepton mass and mixing  in a Neutrino Mass Model  \\ based on $S_4$ flavor symmetry}

\author{V. V. VIEN$^{(1), (2)}$} \address{(1) Institute of Research and Development, Duy Tan University,\\ 182 Nguyen Van Linh, Da Nang City, Vietnam,\\
(2) Department of Physics, Tay
Nguyen University, \\
567 Le Duan, Buon Ma Thuot, DakLak, Vietnam\\
wvienk16@gmail.com}

\maketitle
\begin{history}
\received{Day Month Year}
\revised{Day Month Year}
\end{history}

\begin{abstract}

We study a neutrino mass model based on $S_4$ flavor symmetry which accommodates lepton mass, mixing with non-zero $\theta_{13}$ and CP violation phase. The spontaneous symmetry breaking in the model is imposed to obtain the realistic neutrino mass and mixing pattern at the tree- level with renormalizable interactions. Indeed, the neutrinos get
small masses from one $SU(2)_L$ doubplet and two $SU(2)_L$ singlets  in which one being in $\underline{2}$ and the two others in $\underline{3}$ under $S_4$ with both the breakings $S_{4}\rightarrow S_3$ and $S_{4}\rightarrow Z_3$ are taken place in charged lepton sector and $S_4\rightarrow \mathcal{K}$ in neutrino sector. The model also gives a remarkable prediction of Dirac CP violation $\delta_{CP}=\frac{\pi}{2}$ or $-\frac{\pi}{2}$ in the both normal and inverted spectrum which is still missing in the neutrino mixing matrix. The relation between lepton mixing angles is also represented.


\end{abstract}

\maketitle

\section{\label{intro} Introduction}
The Standard Model (SM) is one of the most successful theories in the elementary particle physics, however, it leaves some unresolved issues that have been empirically verified, such as the fermion masses and mixing,  the mass hierarchies problem and the CP-violating phases. It is obvious that the SM must be extended. Theoretically, there are several proposals for explanation of smallness of neutrino mass and large
 lepton mixing such as the Neutrino Minimal Standard Model \cite{seesaw1,seesaw2, nuMSM1,nuMSM2,nuMSM3,nuMSM4,nuMSM5} , Two-Higgs-doublet model \cite{THDM} , the scotogenic model\footnote {Depending on the particle content, there exist models which generate an
active neutrino mass at 1-loop \cite{1loop} , 2-loop \cite{2loop, 2loop1} , or 3-loop \cite{3loop, 3loop1} level, but Ma's scotogenic model seems to be the
simplest extension.}  \cite{MaSM} , the Georgi-Glashow model \cite{SU5} , $SO(10)$ grand
unification \cite{SO10} , the texture zero models\footnote{For some other scenarios of this type of model,  the reader can see in Ref. \citen{Tzero3} .} \cite{Tzero, Tzero1, Tzero2} , the 3-3-1 models\cite{331s, 331s1, 331s2, 331s3, 331s4, 331s5} and so on. Among the possible extensions of SM, probably the simplest one obtained by adding right-handed neutrinos to its original structure which has been studied in  Refs.  \citen{seesaw1,seesaw2, nuMSM1,nuMSM2,nuMSM3,nuMSM4,nuMSM5}. However, these extensions do not provide a natural explanation for large mass splitting between neutrinos and the lepton mixing was not explicitly explained \cite{TA} . 

There are five well-known patterns of lepton mixing \cite{mixings} ,however, the Tri-bimaximal one proposed by
Harrison-Perkins-Scott (HPS) \cite{hps1, hps2, hps3, hps4}
\be
U_{\mathrm{HPS}}=\left(
\begin{array}{ccc}
\frac{2}{\sqrt{6}}       &\frac{1}{\sqrt{3}}  &0\\
-\frac{1}{\sqrt{6}}      &\frac{1}{\sqrt{3}}  &\frac{1}{\sqrt{2}}\\
-\frac{1}{\sqrt{6}}      &\frac{1}{\sqrt{3}}  &-\frac{1}{\sqrt{2}}
\end{array}\right),\label{Uhps}
\ee
 seems to be the most popular and can be considered as a leading order approximation for the recent
neutrino experimental data. In fact, the absolute
values of the entries of the lepton mixing matrix $U_{PMNS}$ are given in Ref. \citen{Gonzalez2014}
\bea 
\left|U_{\mathrm{PMNS}}\right|=\left(
\begin{array}{ccc}
 0.801 \to 0.845 &\qquad
    0.514 \to 0.580 &\qquad
    0.137 \to 0.158
    \\
    0.225 \to 0.517 &\qquad
    0.441 \to 0.699 &\qquad
    0.614 \to 0.793
    \\
    0.246 \to 0.529 &\qquad
    0.464 \to 0.713 &\qquad
    0.590 \to 0.776
\end{array}\right).\label{Uij}
\eea
The best fit values of neutrino mass squared differences and the leptonic
mixing angles given in Ref. \citen{Gonzalez2014} as shown in Tabs. \ref{NormalH}
and \ref{InvertedH}.  

\begin{table}[h]
\tbl{The experimental values of neutrino mass
squared splittings and leptonic mixing parameters, taken from
Ref. \cite{Gonzalez2014} for normal hierarchy.}
{\begin{tabular}{@{}cccc@{}} \toprule
               &\hs Best fit $\pm1\sigma$  &\hs $3\sigma $ range& \\
\noalign{\smallskip}\hline\noalign{\smallskip}
$\De m_{21}^{2}$[$10^{-5}$eV$^2$] &\hs  $7.50^{+0.19}_{-0.17}$ &\hs  $7.02\to 8.09$ &\\
$\De m_{31}^{2}$[$10^{-3}$eV$^2$]&\hs  $2.457^{+0.047}_{-0.047}$ &\hs  $2.317\to2.607$ &  \\
$\sin ^{2}\theta _{12}$ &\hs   $0.304^{+0.013}_{-0.012}$ &\hs  $0.270 \to 0.344$ &   \\
 $\sin ^{2}\theta _{23}$&\hs   $0.452^{+0.052}_{-0.028}$ &\hs  $0.382 \to 0.643$ & \\
 $\sin ^{2}\theta_{13}$&\hs  $0.0218^{+0.0010}_{-0.0010}$ &\hs  $0.0186 \to 0.0250$ &  \\ 
 $\delta{[^\circ]}$&\hs  $306^{+39}_{-70}$ &\hs  $0 \to 360$ &  \\ \botrule
\end{tabular} \label{NormalH}}
\end{table}
\begin{table}[h]
\tbl{The experimental values of neutrino mass
squared splittings and leptonic mixing parameters, taken from
Ref. \citen{Gonzalez2014} for inverted hierarchy.}
{\begin{tabular}{@{}cccc@{}} \toprule
& Best fit $\pm1\sigma$  & $3\sigma $ range& \\
\noalign{\smallskip}\hline\noalign{\smallskip}
$\De m_{21}^{2}$[$10^{-5}$eV$^2$] &\hs  $7.50^{+0.19}_{-0.17}$ &\hs  $7.02\to 8.09$ &\\
$\De m_{31}^{2}$[$10^{-3}$eV$^2$]&\hs  $-2.449^{+0.048}_{-0.047}$ &\hs  $-2.590\to -2.307$ &  \\
$\sin ^{2}\theta _{12}$ &\hs   $0.304^{+0.013}_{-0.012}$ &\hs  $0.270 \to 0.344$ &   \\
 $\sin ^{2}\theta _{23}$&\hs   $0.579^{+0.025}_{-0.037}$ &\hs  $0.389 \to 0.644$ & \\
 $\sin ^{2}\theta_{13}$&\hs  $0.0219^{+0.0011}_{-0.0010}$ &\hs  $0.0188 \to 0.0251$ &  \\ 
 $\delta{[^\circ]}$&\hs  $254^{+63}_{-62}$ &\hs  $0 \to 360$ &  \\ \botrule
\end{tabular} \label{InvertedH}}
\end{table}

The large lepton mixing angles given in Tabs. \ref{NormalH}, \ref{InvertedH} are completely different from the quark
 mixing ones defined by the Cabibbo- Kobayashi-Maskawa (CKM) matrix \cite{CKM, CKM1} .
 This has stimulated works on flavor symmetries and non-Abelian discrete
 symmetries, which are considered to be the most attractive candidate to
formulate dynamical principles that can lead to the flavor mixing
patterns for quarks and leptons. There are various recent models based
on the non-Abelian discrete symmetries, see for example $A_4$ \cite{A41, A42, A43, A44, A45,A46, A47, A48, A49, A410, A411, A412, A413, A414,
 A415, A416, A417, A418, dlsh} , $S_3$\cite{S31,S32,S33,S34,S35,S36,S37,S38,S39,S310,S311,S312,S313,S314,S315,S316,S317,S318,S319,S320,S322,S323,S324,S325,S326,S327,S328,S329,S330,S331,S332,S333,S334,S335,S336,S337,S338,S339,S340,S341,S342} ,
$S_4$ \cite{S41,S42,S43,S44,
S45,S46,S47,S48,S49,S410,S411,S413,S414,S415,S416,S417,S418,S419,S420,S421,S422,S423,S424,S425,S426,S427,S428,S429, S430} ,
$D_4$ \cite{D41,D42,D43,D45,D46,D47,D48,D49,D410,D411,D412} , \textbf{$T'$ \cite{Tp1,Tp2,Tp3,Tp4,Tp7,Tp8,Tp9,Tp10,Tp11,Tp12} ,} $T_7$
\cite{T71, T72, T73, T74, T75} . However, in all these papers, the fermion masses and mixings generated from non-renormalizable interactions or at loop level but not at tree-level. 

In this work, we investigate another choice with $S_4$ group, the permutation group of four objects, which is also the
symmetry group of a cube. It has 24 elements divided into 5
conjugacy classes, with \underline{1}, \underline{1}$'$,
\underline{2}, \underline{3}, and \underline{3}$'$ as its 5
irreducible representations.  A brief of the theory of $S_4$
group is given in \cite{dlsvS4}. We note that $S_4$ has not been considered before in this kind of the model in this scenario\footnote{In this scenario, fermion masses and mixing angles are generated from renormalizable Yukawa interactions and at tree-level.}. This model is diferent from our previous works \cite{dlsvS4, dlnvS3, vlD4, vlS4, vlS3,vlT7, vD4, vT7, vlkS4,  vlD4q, vlA4, vla27, alv27} because the 3-3-1 models (based on $SU(3)_C\otimes SU(3)_L\otimes U(1)_X$) itself is an extension of the SM.

The rest of this work is organized
as follows. In Sec.
\ref{lepton} we present the necessary elements of the model
and introduce necessary Higgs fields responsible for the
lepton masses. Sec. \ref{quark} is devoted for the quark mass and
mixing at tree level. We summarize our results and make conclusions in the
section \ref{conclus}.  \ref{S4group} briefly provides the theory of $S_4$ group 
with its Clebsch-Gordan coefficients. \ref{S4breaking1}, \ref{S4breaking2} and \ref{S4breaking3} provide the breakings of $S_4$ by $\underline{3}$, $\underline{3}'$ and $\underline{2}$, respectively. 
\section{Lepton mass and mixing\label{lepton}}

The symmetry group of the model under consideration is
\bea
G=\mathrm{SU}(3)_C\otimes\mathrm{SU}(2)_L\otimes \mathrm{U}(1)_Y\otimes \mathrm{U}(1)_X\otimes \underline{S}_4, \label{G}\eea
where the electroweak sector of the SM is supplemented by an auxilliary
symmetry $\mathrm{U}(1)_X$ plus a $S_4$ flavour symmetry  whereas the strong interaction one is retained. The reason for adding the auxiliary
symmetry $\mathrm{U}(1)_X$ was discussed fully in \cite{HeU1X} .
The lepton content of the model, under $[\mathrm{SU}(2)_L,
\mathrm{U}(1)_Y, \mathrm{U}(1)_X,\underline{S}_4]$, is summarized in Tab. \ref{Lepcon}.

\begin{table}[h]
\tbl{\label{Lepcon} The lepton content of the model.}
{\begin{tabular}{@{}ccccccccc@{}} \toprule
   Fields & $\psi_{1,2,3L}$ &$l_{1 (2,3)R}$&\,\,$\nu_{R}$\,\,&\,\,$\phi$\,\,&\,\,$\phi'$\,\,&\,\,$\varphi$\,\,&\,\,$\chi$\,\,&\,\,$\zeta$ \\
\noalign{\smallskip}\hline\noalign{\smallskip}
$\mathrm{SU}(2)_L$  & $2$ &$1$&$1$&   $2$  &    $2$   &    $2$     &$1$ & $1$   \\
$\mathrm{U}(1)_Y$  & $-1$ &$-2$&$0$&   $1$  & $1$    & $1$&$0$&$0$   \\ 
$\mathrm{U}(1)_X$ & $1$  &$1$&$0$&   $0$  & $0$  & $-1$ &$0$ & $0$  \\
$\underline{S}_4$&  $\underline{3}$  &$\underline{1} (\underline{2})$&$\underline{3}$&   $\underline{3}$  & $\underline{3}'$  & $\underline{1}$ & $\underline{3}$&$\underline{2}$  \\ \botrule
\end{tabular} }
\end{table}

The charged lepton masses arise from the couplings of
$\bar{\psi}_{L} l_{1R}$ and $\bar{\psi}_{L} l_{R}$ to scalars, where $\bar{\psi}_{L} l_{1R}$
transforms as $2$ under $\mathrm{SU}(2)_L$ and $\underline{3}$
under $S_4$; $\bar{\psi}_{L} l_{R}$ transforms as $2$ under $\mathrm{SU}(2)_L$ and $\underline{3}\oplus \underline{3}'$ under $S_4$.
To generate masses for the charged leptons, we need two scalar
multiplets $\phi$ and $\phi'$ given in Tab. \ref{Lepcon}.

The Yukawa interactions are
 \bea -\mathcal{L}_{l}&=&h_1 (\bar{\psi}_{L}\phi)_{\underline{1}} l_{1R}+
 h_2 (\bar{\psi}_{L}\phi)_{\underline{2}}l_{R}
+h_3 (\bar{\psi}_{i L}\phi')_{\underline{2}} l_{R}+H.c.\label{Lclep0}\eea
Theoretically, a possibility that the Tribimaximal mixing matrix ($U_{HPS}$) can be decomposed into only two independent rotations
may provide a hint for some underlying structure in the lepton sector,
\bea
U_{\mathrm{HPS}}&=&\left(
\begin{array}{ccc}
\frac{2}{\sqrt{6}}       &\frac{1}{\sqrt{3}}  &0\\
-\frac{1}{\sqrt{6}}      &\frac{1}{\sqrt{3}}  &-\frac{1}{\sqrt{2}}\\
-\frac{1}{\sqrt{6}}      &\frac{1}{\sqrt{3}}  &\frac{1}{\sqrt{2}}
\end{array}\right)
=\frac{1}{\sqrt{3}}\left(
\begin{array}{ccc}
1       &1  &1\\
1     &\om^2  &\om\\
1    &\om  &\om^2 \end{array}\right)\left(
\begin{array}{ccc}
0      &1 &0\\
\frac{1}{\sqrt{2}}      &0 &-\frac{i}{\sqrt{2}}\\
\frac{1}{\sqrt{2}}      &0 &\frac{i}{\sqrt{2}}
\end{array}\right)\cong U^+_L U_{\nu},\label{Ulepdecomp}
\eea
where $\omega = \exp(2 \pi i/3) = -1/2 + i\sqrt{3}/2$.

All possible breakings of $S_4$ group under triplets $\underline{3}$ and $\underline{3}'$ are presented in appendices \ref{S4breaking1} and \ref{S4breaking2}, respectively.  
To obtain charged - lepton mixing satisfying (\ref{Ulepdecomp}), in this work we argue that both the breakings 
$S_{4}\rightarrow S_3$ and $S_{4}\rightarrow Z_3$ are taken place in charged lepton sector. The breaking $S_{4}\rightarrow S_3$ 
can be achieved by a $SU(2)_L$ doublet $\phi$ with the third alignment given in \ref{S4breaking1}, i.e, $\langle \phi\rangle=(\langle \phi_1\rangle,
\langle \phi_1\rangle,\langle \phi_1\rangle )$ under $S_4$, where
\be
\langle \phi_1\rangle=(0\hs v )^T, \label{vevphi}
\ee
and the breaking $S_{4}\rightarrow Z_3$ can be achieved by another $SU(2)_L$ doublet $\phi'$ with the third alignment given in \ref{S4breaking2}, i.e, $\langle \phi'\rangle=(\langle \phi'_1\rangle,\langle \phi'_1\rangle,\langle \phi'_1\rangle )$ under $S_4$, where
\be
\langle \phi'_1\rangle=(0\hs v' )^T. \label{vevphip}
\ee
After electroweak breaking, the mass Lagrangian for the charged leptons becomes
\bea
-\mathcal{L}^{\mathrm{mass}}_l
&=&(\bar{l}_{1L},\bar{l}_{2L},\bar{l}_{3L})
M_l (l_{1R},l_{2R},l_{3R})^T+H.c,\label{Lclep2}\eea
where \be M_l=
\left(%
\begin{array}{ccc}
  h_1v & h_2v-h_3v' & h_2v+h_3v' \\
  h_1v & \hs (h_2v-h_3v')\om & \hs\, (h_2v+h_3v')\om^2 \\
  h_1v & \hs\,\, (h_2v-h_3v')\om^2 & \,\,\,\,\, (h_2v+h_3v')\om \\
\end{array}%
\right).\label{Ml}\ee
The mass matrix $M_l$ in Eq. (\ref{Ml}) is diagonalized by $U^\dagger_L M_lU_R =diag (m_e, \, m_\mu,\, m_\tau)$,  with   
\bea m_e &=&\sqrt{3}h_1 v, \hs m_\mu=\sqrt{3}(h_2 v - h_3v'), \hs m_\tau=\sqrt{3}(h_2 v+h_3v') ,\label{clepmass}\eea 
and\footnote{The charged lepton mixing matrix in this model given in Eq. (\ref{Uclep})  is the same as that of in Refs.\citen{vT7, vlS4, vlA4} and a little different from that in Ref.\citen{vlT7}.}  \bea U_L=\fr{1}{\sqrt{3}}\left(%
\begin{array}{ccc}
  1 & 1 & 1 \\
  1 & \om & \om^2 \\
  1 & \om^2 & \om \\
\end{array}%
\right),\hs U_R=1.\label{Uclep}\eea The result in Eq. (\ref{clepmass}) shows that the masses of muon
and tauon are separated by the $SU(2)_L$ doublet $\phi'$. This is the reason
why $\phi'$  was additional introduced to $\phi$ in lepton sector.\\
Now, by combining Eq. (\ref{clepmass}) with the experimental values for masses of  the charged leptons given in Ref. \citen{PDG2014},
\be m_e\simeq0.51099\, \textrm{MeV},\hs \ m_{\mu}=105.65837 \ \textrm{MeV},\hs m_{\tau}=1776.82\,
\textrm{MeV} \label{Lepmas}\ee
It follows that $h_1\ll h_2, h_3$ and $h_2\simeq h_3$ if $v'\simeq v$. On the other hand, if we suppose that \footnote{In the SM, the
Higgs VEV $v$ is 246 GeV, fixed by the $W$ boson mass and the gauge coupling, $m^2_W=\frac{g^2 }{4}v^2_{weak}$. However, in the model under consideration,  $M^2_W\simeq\frac{g^2}{2}\left(3v^2+3v'^2\right)$). Therefore, we can identify 
$v^2_{weak}=6(v^2+v'^2) = (246 \, \mathrm{GeV})^2$ and then obtain $v'\simeq v\simeq 71 \, \mathrm{GeV}$. In this work, we chose
 $v= 100 \, \mathrm{GeV}$ for its scale.} $v \sim 100\, \mathrm{GeV}$ then
\bea
h_1\sim 10^{-6},\,\,\, h_2\sim h_3\sim 10^{-3},\label{hi}\eea
i.e, in
the model under consideration, the hierarchy between the masses for charged-leptons can be achieved if there exists a hierarchy between 
Yukawa couplings $h_1$ and $h_{2,3}$ in charged-lepton sector as given in Eq. (\ref{hi}).

The neutrino masses arise from the couplings of $\bar{\psi}_{L} \nu_{R}$
and $\bar{\nu}^c_{R} \nu_{R}$ to scalars, where $\bar{\psi}_{L} \nu_{R}$ transforms as
$2$ under $\mathrm{SU}(2)_L$ and $\underline{1}\oplus\underline{2}\oplus\underline{3}_s\oplus
\underline{3}'_a$ under $S_4$;
$\bar{\nu}^c_{R} \nu_{R}$ transform as $1$ under $\mathrm{SU}(2)_L$
and $\underline{1}\oplus\underline{2}\oplus\underline{3}_s\oplus
\underline{3}'_a$ under $S_4$. Note that under $S_4$ symmetry, each tensor product $\underline{3} \otimes \underline{3} \otimes \underline{3}$  contains one invariant\footnote{In fact $\underline{3} \otimes \underline{3}'\otimes \underline{3}$ has one invariant but this invariant vanishes in neutrino sector since $(\underline{3} \otimes \underline{3}')_{\underline{3}_a}$ containts $\underline{3}_a(23-32, 31-13, 12-21)$ under $S_4$.}.
On the other hand, $2\otimes 2=1\oplus 3$ and
$3\otimes 3=1\oplus 3\oplus 5$ under $SU(2)_L$.
For the known
$SU(2)_L$ scalar doublets, only two available interactions $(\bar{\psi}_L \tilde{\phi})_{\underline{3}_s}\nu_{R}, (\bar{\psi}_L \tilde{\phi}')_{\underline{3}_a}\nu_{R}$, but explicitly suppressed because of the
$U(1)_X$ symmetry. We therefore additionally introduce
one $SU(2)_L$ doublet $(\varphi)$ and two $SU(2)_L$ singlets $(\chi,\, \zeta)$ , respectively, put in $\underline{1}$, $\underline{3}$ and $\underline{2}$ under $S_4$ as given in Tab. \ref{Lepcon}.

It is need to note that $\varphi$ contributes to the Dirac mass matrix in
the neutrino sector and $\chi$ contributes to the Majorana mass matrix of the right-handed neutrinos. We also note that the $U(1)_X$ symmetry
 forbids the Yukawa terms of the form $(\bar{\psi}_L \tilde{\phi})_{\underline{3}_s}\nu_{R}$ and yield the expected results in neutrino sector, and this is interesting feature of $X$-symmetry.

All possible breakings of $S_4$ group under triplet $\underline{3}$ and doublet $\underline{2}$ are given in appendices \ref{S4breaking1} and \ref{S4breaking3}, respectively. To obtain a realistic neutrino spectrum, i.e, resulting the non-zero $\theta_{13}$ and CP violation, in this work, we argue that the breaking
$S_4\rightarrow \mathcal{K}$ must be taken place in neutrino sector. This can be achieved within each
case below. 
\ben
\item A $\mathrm{SU}(2)_L$ doublet $\chi$ put in $\underline{3}$ under $S_4$  with the VEV is chosen by
\bea
\langle \chi_1\rangle&=&v_\chi, \hs \langle \chi_2\rangle=\langle \chi_3\rangle=0. \label{vevchi}
\eea

\item Another  $\mathrm{SU}(2)_L$ doublet $\zeta$ put in $\underline{2}$ under $S_4$ with the VEV given by
\bea
\langle \zeta\rangle &=& (\langle \zeta_1\rangle, \langle \zeta_2\rangle  ),\hs
\langle \zeta_i\rangle = v_{\zeta_i} \hs (i=1,2) . \label{vevzeta}\eea
\een

The Yukawa Lagrangian invariant under $G$ symmetry in neutrino sector reads:
\bea
 -\mathcal{L}_{\nu}&=&\frac{x}{2} (\bar{\psi}_L \tilde{\varphi})_{\underline{3}}\nu_{R}
+\frac{y}{2} (\bar{\nu}^c_R\chi)_{\underline{3}_s}\nu_{R}
+\frac{M}{2} \bar{\nu}^c_R\nu_{R}+\frac{z}{2} (\bar{\nu}^c_R\zeta)_{\underline{3}}\nu_{R}+H.c,\label{Lny}\eea
where $M$ is the bare Majorana mass for the right-handed neutrino.

After electroweak breaking, the mass Lagrangian for the
neutrinos is given by
\bea -\mathcal{L}^{mass}_{\nu}&=&\fr 1 2
\bar{\chi}^c_L M_\nu \chi_L+ H.c.,\label{nm}\eea where \bea
\chi_L&\equiv& \left(\nu_L \hs
  \nu^c_R \right)^T,\hs\,\,\, M_\nu\equiv\left(%
\begin{array}{cc}
  0 & M_D \\
  M_D & M_R \\
\end{array}%
\right), \label{MnuLDR}\\
 \nu_L&=&(\nu_{1L}\hs\nu_{2L}\hs\nu_{3L})^T,\hs
\nu^c_R=(\nu^c_{1R}\hs \nu^c_{2R}\hs\nu^c_{3R})^T, \nn  \eea and the
mass matrices $M_{D}, M_{R}$ are
then obtained by
\bea M_D&=&
m_D\left(%
\begin{array}{ccc}
 1 & 0 & 0 \\
 0 &1 & 0 \\
 0&0 & 1 \\
\end{array}%
\right),\, M_R=
\left(%
\begin{array}{ccc}
M+M_1+M_2 & 0 & 0 \\
 0 &M+\om M_1+\om^2 M_2 & M' \\
 0 &M' & M+\om^2 M_1+\om M_2 \\
\end{array}%
\right), \crn
 M'&=&yv_{\chi},\,\, m_D=x v_{\varphi},  \,\, M_i= z v_{\zeta_i}\,\, (i=1,2) ,  \label{MDR}\eea 
with $v_\varphi = \langle \varphi \rangle$, and $M_D$ is the Dirac neutrino mass
matrix, $M_R$ is the right-handed Majorana neutrino mass matrix.

The effective neutrino mass matrix, in the framework of seesaw mechanism, is given by
 \bea
{M}_{\mathrm{eff}}&=&-M_D^T{M}_R^{-1}M_D=\left(%
\begin{array}{ccc}
  A & 0 & 0 \\
  0 & B_1 & C \\
  0 & C & B_2 \\
\end{array}%
\right), \label{Meff}\eea
where  
\bea
A&=&-\frac{m^2_D}{M+M_1+M_2},\hs B_{1,2}=\frac{m^2_D\left[-2M+M_1+M_2\pm i\sqrt{3}(M_1-M_2)\right]}{2\mathfrak{M}},\crn C&=&\frac{m^2_DM'}{\mathfrak{M}}, \hs
\mathfrak{M}=M^2+M^2_1+M^2_2-MM_1-MM_2-M_1M_2-M'^2.\label{ABCM}
\eea
The matrix $M_{\mathrm{eff}}$ in (\ref{Meff}) can be diagonalized as follows $U^T_\nu
 M_{\mathrm{eff}} U_\nu=\mathrm{diag}(m_1,  m_2,  m_3)$, with 
\bea m_1
&=&\fr 1 2 \left(B_1 + B_2 + \sqrt{(B_1 + B_2)^2+4C^2}\right),\hs 
m_2=A,\crn
m_3&=&\fr 1 2 \left(B_1 + B_2 - \sqrt{(B_1 + B_2)^2+4C^2}\right),\label{m123}\eea
and \bea U_\nu&=&\left(%
\begin{array}{ccc}
  0 & 1 & 0 \\
  \fr{1}{\sqrt{K^2+1}} & 0 & \fr{K}{\sqrt{K^2+1}} \\
  -\fr{K}{\sqrt{K^2+1}} & 0 & \fr{1}{\sqrt{K^2+1}} \\
\end{array}%
\right).\left(%
\begin{array}{ccc}
  1 & 0 & 0 \\
 0 & 1 & 0 \\
 0 & 0& i \\
\end{array}%
\right),\label{Unu1}\\
K&=&\frac{B_1 -B_2 -\sqrt{(B_1-B_2)^2+4 C^2}}{2C}.\label{K}
\eea
The lepton mixing matrix, obtained from the matrices $U_\nu$ and $U_L$ in Eqs. (\ref{Uclep}) and (\ref{Unu1}), is expressed as
\bea U=U^\dagger_L
U_\nu= \fr{1}{\sqrt{3}}\left(%
\begin{array}{ccc}
  \fr{1-K}{\sqrt{K^2+1}} & 1 &  \fr{1+K}{\sqrt{K^2+1}} \\
\fr{\om(\om-K)}{\sqrt{K^2+1}} & 1 &  \fr{\om(1+K\om)}{\sqrt{K^2+1}} \\
\fr{\om(1-K\om)}{\sqrt{K^2+1}} & 1 &  \fr{\om(\om+K)}{\sqrt{K^2+1}} \\
\end{array}%
\right).\left(%
\begin{array}{ccc}
  1 & 0 & 0 \\
 0 & 1 & 0 \\
 0 & 0& i \\
\end{array}%
\right).\label{Ulep}\eea
where $K$ is defined in Eq.(\ref{K}). 

In the standard parametrization, the lepton mixing
 matrix can be parametrized as \cite{PDG2014}
\be
       U_{PMNS} = \begin{pmatrix}
    c_{12} c_{13}     & -s_{12} c_{13}                    & - s_{13} e^{-i \delta}\\
    s_{12} c_{23}-c_{12} s_{23} s_{13}e^{i \delta} & c_{12} c_{23}+s_{12} s_{23} s_{13} e^{i \delta} & - s_{23} c_{13}\\
    s_{12} s_{23}+c_{12} c_{23} s_{13}e^{i \delta}& c_{12} s_{23}-s_{12} c_{23} s_{13} e^{i \delta}  &\,\,  c_{23} c_{13} \\
     \end{pmatrix} \times P, \label{Ulepg}
\ee 
where $P=\mathrm{diag}(1, e^{i \alpha}, e^{i \beta})$, and
$c_{ij}=\cos \theta_{ij}$, $s_{ij}=\sin \theta_{ij}$ with
$\theta_{12}$, $\theta_{23}$ and $\theta_{13}$ being the
solar, atmospheric and  reactor angles, respectively, and $\delta= [0, 2\pi]$ is the Dirac
 CP violation phase while $\alpha$ and
$\beta$ are two Majorana CP violation phases.

Comparing the lepton mixing matrix in Eq. (\ref{Ulep}) to the standard parametrization in Eq.(\ref{Ulepg}), one obtains
$\alpha=0,  \beta =\pi/2$, and 
 \bea s_{13} e^{-i \delta}&=&\fr{-1-K}{\sqrt{3}\sqrt{K^2+1}},\label{s13}\\
 t_{12}&=&\frac{\sqrt{K^2+1}}{K-1},\label{t12}\\
  t_{23}&=&-\frac{1+K\om}{K+\om}.\label{t23}\eea
Substituting $\om =-\frac{1}{2}+i\frac{\sqrt{3}}{2}$ into Eq. (\ref{t23}) yields: \bea 
Re K&=&\frac{t^2_{23}-4t_{23}+1}{2(t^2_{23}-t_{23}+1)},\hs
Im K=\frac{\sqrt{3}}{2}\frac{1-t^2_{23}}{t^2_{23}-t_{23}+1}.\label{imreK}
\eea 
It is easily to see that $|K|=\sqrt{(Im K)^2+(Re K)^2}=1$. 
Combining Eq. (\ref{s13}) and Eq. (\ref{t12})  we obtain: \bea
e^{-i\delta}&=&\frac{1}{\sqrt{3}s_{13}t_{12}}\frac{1+K}{1-K}.\nn\eea
or \bea
-i\frac{t_{23}-1}{s_{13}t_{12}(t_{23}+1)}
=\cos\delta-i\sin\delta. \label{add}\eea 
By equating the real and imaginary parts of the equation (\ref{add}), we get\bea \cos\delta&=&0,\hs
\sin\delta
=\frac{t_{23}-1}{s_{13}t_{12}(t_{23}+1)}.\label{sindel} \eea
Since $\cos\delta=0$ so that $\sin\delta$ must be equal to $\pm1$,
it is then $\delta =\frac{\pi}{2}$ or $\delta =-\frac{\pi}{2}$.
The value of the Jarlskog invariant
$J_{CP}$ which determines the magnitude of CP violation in
neutrino oscillations is determined  \cite{PDG2014} \be
J_{CP}=\frac{1}{8}\cos\theta_{13}\sin2\theta_{12}\sin2\theta_{23}\sin2
\theta_{13}\sin\delta.\label{J1}
\ee
Once $\theta_{12}$, $\theta_{23}$ and $\theta_{13}$ have been
determined experimentally, the size of
$J_{CP}$ depends essentially only on the magnitude of the currently not well determined value of the Dirac
phase $\delta$.
 Thus, our model predicts the maximal Dirac CP violating phase which
  is the same as in Refs. \citen{TBM2, MaximalCP} but the difference comes from $\theta_{23}$. Namely, in Refs. \citen{TBM2, MaximalCP} $\theta_{23}= \pi/4$ but in our model $\theta_{23}\neq \pi/4$ which is more consistent with the recent experimental data given in Tabs. \ref{NormalH}, \ref{InvertedH} and this is one of the most striking prediction
  of the model under consideration.

At present, the precise evaluation of $\theta_{23}$ is still an open
problem while $\theta_{12}$ and $\theta_{13}$ are now very
constrained \cite{PDG2014} . From Eq. (\ref{sindel}), as will see below, our model can provide constraints on
$\theta_{23}$ from $\theta_{12}$ and $\theta_{13}$ which satisfy the data given in Ref.\citen{PDG2014}.

\begin{itemize}
\item[(i)]  In the case $\delta = \frac{\pi}{2}$, from (\ref{sindel}) we have the relation
 among three Euler's angles as follows: \bea
t_{23}&=&\frac{1+s_{13}t_{12}}{1-s_{13}t_{12}},
\label{relat1} \eea
or
\bea
s^2_{23}&=&\frac{(1-s^2_{12})\left(1+\sqrt{\frac{s^2_{12}s^2_{13}}{1-s^2_{12}}}\right)^2}{2[1+s^2_{12}(s^2_{13}-1)]}.\label{relat2} \eea

In Fig. \ref{relat1v}, we have plotted the
values of $s^2_{23}$ as a function of
 $s^2_{12}$ and $s^2_{13}$ with $s^2_{12}\in (0.270, 0.344)$, $s^2_{13} \in (0.0186, 0.0250)$  given in Ref. \citen{Gonzalez2014} in the case $\delta =\frac{\pi}{2}$ at the $3\sigma$ level.

\begin{figure}[h]
\begin{center}
\includegraphics[width=8.0cm, height=6.0cm]{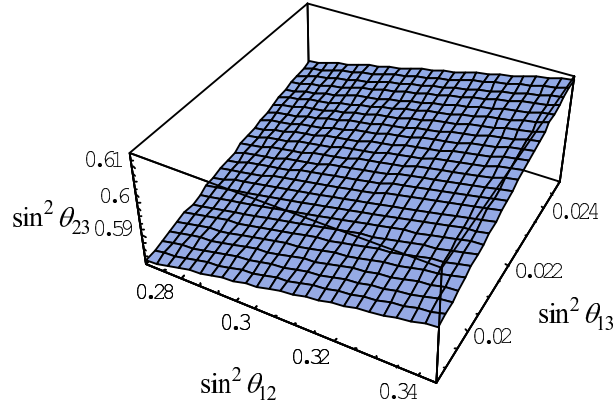}
\vspace*{-0.4cm} \caption[$s^2_{23}$ as a function of
 $s^2_{12}$ and $s^2_{13}$ with $s^2_{12}\in (0.270, 0.344)$, $s^2_{13} \in (0.0186, 0.0250)$ in the case $\delta =\frac{\pi}{2}$ at the $3\sigma$ level]{$s^2_{23}$ as a function of
 $s^2_{12}$ and $s^2_{13}$ with $s^2_{12}\in (0.270, 0.344)$, $s^2_{13} \in (0.0186, 0.0250)$ in the case $\delta =\frac{\pi}{2}$ at the $3\sigma$ level.}\label{relat1v}
\vspace*{-0.3cm}
\end{center}
\end{figure}
Taking the new data $s^2_{12} =0.30\, (\theta_{12}=33.46^o)$ and $s^2_{13} =0.0245\, (\theta_{13}=9.00^o)$  we obtain
 $s^2_{23}=0.6014$ ,i.e, $\theta_{23}=50.8507^o$ which is larger than $45^o$, and \bea
K=-0.938924 - 0.344125 i, \hs (|K|=1). \label{Kvalues}\eea
The lepton mixing matrix in (\ref{Ulep}) then takes the form
\bea U\simeq\left(%
\begin{array}{ccc}
 0.82841 &\hs 0.57735 &\hs -0.147252\\
-0.53546 &\hs 0.57735 &\hs -0.78743 \\
-0.29295 &\hs 0.57735 &\hs  0.64742 \\
\end{array}%
\right),\label{Ulepmix1}\eea
which is consistent with constraint in Eq. (\ref{Uij}).

Combining (\ref{K}) and the values of $K$ in
(\ref{Kvalues}), we obtain the relation \bea B_1=B_2-(2.75481\times
10^{-7}+0.68825i)C. \label{B1B2v1} \eea 

\item[(ii)]
Similar to the case with $\delta =\frac{\pi}{2}$, in the case $\delta =-\frac{\pi}{2}$, we find the followings relation: \bea
s^2_{23}&=&\frac{(1-s^2_{12})\left(-1+\sqrt{\frac{s^2_{12}s^2_{13}}{1-s^2_{12}}}\right)^2}{2[1+s^2_{12}(s^2_{13}-1)]}. \label{relat2} \eea
In Fig. \ref{relat2v}, we have plotted the
values of $s^2_{23}$ as a function of
 $s^2_{12}$ and $s^2_{13}$ with $s^2_{12}\in (0.270, 0.344)$, $s^2_{13} \in (0.0186, 0.0250)$  given in Ref. \citen{Gonzalez2014} in the case $\delta =-\frac{\pi}{2}$ at the $3\sigma$ level. 

\begin{figure}[ht]
\begin{center}
\includegraphics[width=8.0cm, height=6.0cm]{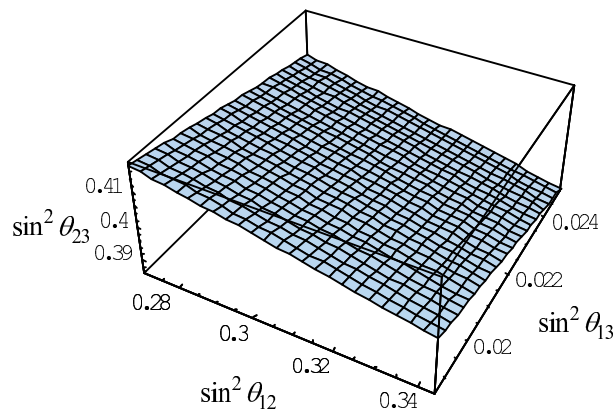}
\vspace*{-0.4cm} \caption[$s^2_{23}$ as a function of
 $s^2_{12}$ and $s^2_{13}$ with $s^2_{12}\in (0.270, 0.344)$, $s^2_{13} \in (0.0186, 0.0250)$ in the case $\delta =-\frac{\pi}{2}$ at the $3\sigma$ level]{$s^2_{23}$ as a function of
 $s^2_{12}$ and $s^2_{13}$ with $s^2_{12}\in (0.270, 0.344)$, $s^2_{13} \in (0.0186, 0.0250)$ in the case $\delta =-\frac{\pi}{2}$ at the $3\sigma$ level.}\label{relat2v}
\vspace*{-0.3cm}
\end{center}
\end{figure}
If $s^2_{12} =0.30$ and $s^2_{13} =0.0245$  we obtain
 $s^2_{23}=0.39860 \, (\theta_{23}=39.15^o)$, and 
\bea
K=-0.938924 + 0.344125 i, \hs (|K|=1). \label{Kvalues1}\eea 
In this case the lepton mixing matrix in (\ref{Ulep})  takes the form:
\bea U\simeq\left(%
\begin{array}{ccc}
0.82967 &\hs 0.57735 &\hs-0.14725 \\
-0.28731 &\hs 0.57735 &\hs -0.64489 \\
-0.54236 &\hs 0.57735 &\hs 0.79214 \\
\end{array}%
\right),\label{Ulepmix1}\eea
The relation between $B_{1,2}$ and $C$ is determined as follows \bea B_1=B_2-(2.75481\times
10^{-7}-0.68825i)C. \label{B1B2v2} \eea 
\end{itemize}
 
\subsection{Normal case ($\Delta m^2_{23}> 0$)}

\begin{figure}[h]
\bc
\includegraphics[width=12.0cm, height=5.0cm]{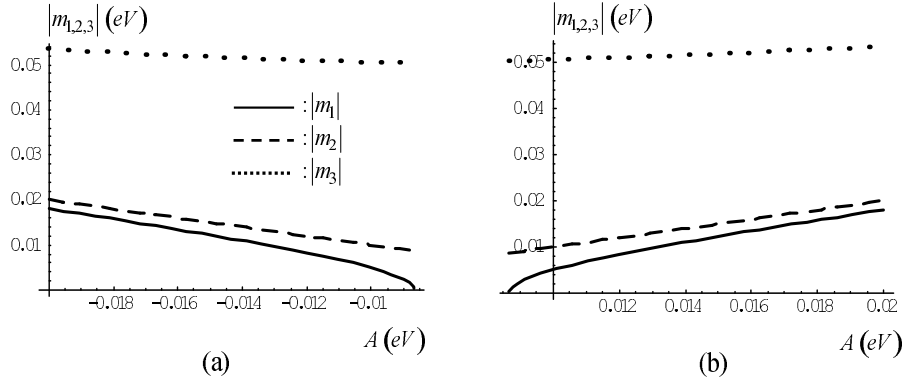}
\vspace*{-0.1cm} \caption[$|m_{1,2,3}|$ as functions of $A$ in
 the normal hierarchy with a) $A\in(-0.02, -0.00867) \, \mathrm{eV}$ and b) $A\in(0.00867, 0.02) \, \mathrm{eV}$.]{$|m_{1,2,3}|$ as functions of $A$ in
 the normal hierarchy with a) $A\in(-0.02, -0.00867) \, \mathrm{eV}$ and b) $A\in(0.00867, 0.02) \,  \mathrm{eV}$.}\label{m123Ncase1}
\ec
\end{figure}

In this case, substituting $B_1$ from
(\ref{B1B2v1}) into (\ref{m123}) and taking the two
experimental data on squared mass differences of neutrinos
given in Ref. \citen{Gonzalez2014}, $\Delta m^2_{21}=7.50\times 10^{-5}\, \mathrm{eV^2}$ and $\Delta m^2_{31}=2.457\times 10^{-3}\, \mathrm{eV^2}$, we get a solution\footnote{The system of equations has two solutions but they have the same absolute values of $m_{1,2, 3}$, the unique
difference is the sign of them. So, here we only consider in
detail the case in Eq. (\ref{B2CN1}).}  (in [eV]) as shown in Appendix \ref{Npi2}.  Using the upper bound on the absolute value of
neutrino  mass Refs. \citen{Tegmark, planck, JCAP15} we can
 restrict the values of $A$,  $A \leq 0.6\,\mathrm{eV}$. However, in the case in (\ref{B2CN1}),
  $\left|A\right| \in (0.00867, 0.02)\, \mathrm{eV}$ can reach the normal neutrino mass hierarchy which is dipicted in Fig. \ref{m123Ncase1}\footnote{The expressions (\ref{B2CN1}) , (\ref{m123}) and (\ref{B1B2v1}) show that $m_{i}\, (i=1,2,3)$
depends only on one parameter $A\equiv m_2$ so we consider $m_{1,3}$ as
functions of $m_2$. However,  to have an explicit hierarchy on
neutrino masses $m_2$ should be
included in the figures.} .  

\begin{figure}[h]
\begin{center}
\includegraphics[width=12.5cm, height=5.5cm]{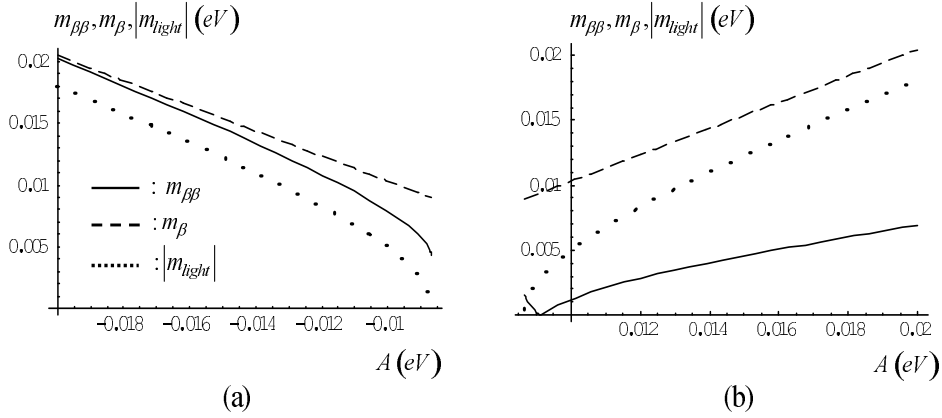}
\caption[$m_{\beta}$, $m_{\beta\beta}$ and $|m_{light}|$ as functions of $A$ in the normal hierarchy with a)
$A \in (-0.02, -0.00867)\,\mathrm{eV}$  and b) $A \in (0.0087, 0.02)\,\mathrm{eV}$ .]{$m_{\beta}$, $m_{\beta\beta}$ and $|m_{light}|$ as functions of $A$ in the normal hierarchy with a)
$A \in (-0.02, -0.00867)\,\mathrm{eV}$  and b) $A \in (0.0087, 0.02)\,\mathrm{eV}$ .}\label{mbbNcase1}
\end{center}
\end{figure}

In the model under consideration, the effective neutrino mass from tritium beta decay $m_\beta = \sqrt{\sum_{i=1}^3 |U_{ei}|^2 m_i^2}$ and the neutrino mass obtained from neutrinoless double-beta decays $m_{\beta \beta} = |\sum_{i=1}^3 U_{ei}^2 m_i|$ are dipicted in Fig. \ref{mbbNcase1}. 
We also note that in the normal spectrum, $|m_1|\approx |m_2|<|m_3|$, so $m_{1}\equiv m_{light}$ is the lightest neutrino mass. 

To get explicit values of the model
parameters, we assume $A=10^{-2}\, \mathrm{eV}$, which is safely small\footnote{The precise value of the mass of neutrinos is still an open question, however, it lies in the range of a few eV.}.
Then the other neutrino masses are explicitly
 given as 
\bea
&& m_1=-5.00\times 10^{-3}\, \mathrm{eV},\, m_2= 10^{-2}\, \mathrm{eV},\,
  m_3\simeq -4.982 \times 10^{-2} \, \mathrm{eV},\label{m123Nor}\\
&&m_{\beta\beta}= 1.88866  \times 10^{-3} \, \mathrm{eV},\, m_{\beta}=1.02156 \times 10^{-2} \, \mathrm{eV},\\
&&|m_1|+|m_2|+|m_3|=6.48197 \times 10^{-2}\, \mathrm{eV},\eea
and
\bea
 B_{1,2}&=&-(2.74098\pm0.821343i)\times 10^{-2} \,\mathrm{eV},\crn 
C&=& (2.38676-1.28329i)\times 10^{-2}\,\mathrm{eV}\simeq 2.38676\times 10^{-2}\,\mathrm{eV}.\label{B12Ccase1N}\eea
 Furthermore, combining Eqs. (\ref{ABCM}) and (\ref{B12Ccase1N}) we get a solution\footnote{This system of equations
  has two solutions, however, these solutions differ only by the sign of $m_D$ (or the sign
of $m_{1,2,3}$) which has no effect on the neutrino oscillation experiments.}:
\bea
M' &=& (2.39395-1.28716\times 10^{-7}i)M, \,\, m_D=(-0.158066 +3.21117\times 10^{-17}i)\sqrt{M},\crn
M_{1,2} &=& (-2.22488 \pm 2.15951\times 10^{-7}i)M. \label{M12MpmD}\eea
and
\bea
x&=&(-0.158066 +3.21117\times 10^{-17}i)\sqrt{M}/v_\varphi\simeq-0.158066\sqrt{M}/v_\varphi,\crn
y&=&  (2.39395-1.28716\times 10^{-7}i)M/v_\chi\simeq 2.39395M/v_\chi ,\crn
z&=&(-2.22488 + 2.15951\times 10^{-7}i)M/v_{\zeta_1}\simeq -2.22488 M/v_1,\label{xyz}\\
v_{\zeta_2}&=&(0.572442+1.52625\times 10^{-7}v_{\zeta_1})\simeq 0.572442 v_{\zeta_1}.\label{v1v2}
\eea
Eq. (\ref{v1v2}) shows that $v_{\zeta_1}$ and $v_{\zeta_2}$ are different  from each other but in the same order of 
magnitude \footnote{In the case $v_{\zeta_1}= v_{\zeta_2}$, i.e, $M_1=M_2$, the lepton mixing matrix $U_{Lep}$ in Eq. (\ref{Ulep})
 becomes an exact Tri-bimaximal mixing which can be considered as a good approximation for the recent neutrino experimental data. Hence, the condition $v_{\zeta_1}\neq v_{\zeta_2}$ is necessary to reach the realistic neutrino spectrum, and the relation (\ref{v1v2}) is satisfy this condition.}.
The solution in Eq. (\ref{m123Nor}) constitutes the normal spectrum and consistent with
the constraints on the absolute value of the neutrino masses \cite{Gonzalez2014, PDG2014, JCAP15}.

\hs  Similarly, in the case $\delta =-\frac{\pi}{2}$, the numerical fit of all parameters to lepton mass and
mixing data is summarized in Tab. \ref{tb2N}.

\begin{table}[h]
\tbl{\label{tb2N} The observables and parameters of the model in the case $\delta =-\pi/2$.}
{\begin{tabular}{@{}ccccccccc@{}} \toprule
  Observables&Data fit $3\sigma$ range from Ref. \citen{Gonzalez2014} & The values of the model parameters \\
\noalign{\smallskip}\hline\noalign{\smallskip}
$\theta_{12}(^\circ)$  & $31.29 \to 35.91$ &$33.46$    \\
$\theta_{23}(^\circ)$  & $38.2 \to 53.3$ &$39.15$    \\ 
$\theta_{13}(^\circ)$  & $7.87 \to 9.11$ &$9.0$    \\
$\Delta m^2_{21}$  & $(7.02 \to 8.09)\times 10^{-5} \, \mathrm{eV}^2$ &$7.50$    \\
$\Delta m^2_{31}$  & $(2.317 \to 2.607)\times 10^{-3} \,\mathrm{eV}^2$ &$2.457$    \\
$|m_{1}| \,[\mathrm{eV}]$  & $-$ &$5\times 10^{-3}$    \\
$|m_{2}| \,[\mathrm{eV}]$  & $-$ &$10^{-2}$    \\
$|m_{3}| \,[\mathrm{eV}]$  & $-$ &$5.05668 \times 10^{-2}$    \\
$\sum m_{i} \,[\mathrm{eV}]$  & $-$ &$4.55668\times 10^{-2}$    \\
$m_{\beta\beta} \,[\mathrm{eV}]$  & $-$ &$1.20486\times 10^{-3}$    \\
$m_{\beta} \,[\mathrm{eV}]$  & $-$ &$1.02949\times 10^{-3}$    \\
$A \,[\mathrm{eV}]$  & $-$ &$10^{-2}$    \\
$B_{1,2} \,[\mathrm{eV}]$  & $-$ &$(-2.77834 \pm 0.835034 i)\times 10^{-2}$    \\
$C \,[\mathrm{eV}]$  & $-$ &$2.42654\times 10^{-2}$    \\\botrule
\end{tabular} }
\end{table}
 The parameters $x, y, z$ are given as follows:
\bea
x&\simeq&-0.158066\sqrt{M}/v_\varphi,\hs 
y\simeq 2.40384 M/v_\chi ,\hs 
z\simeq 1.27474 M/v_1,\label{xyzN2}\\
v_{\zeta_2}&\simeq& 1.74932 v_{\zeta_1}.
\eea

\subsection{Inverted case ($\Delta m^2_{32}< 0$)}
\begin{figure}[h]
\bc
\includegraphics[width=12.0cm, height=5.0cm]{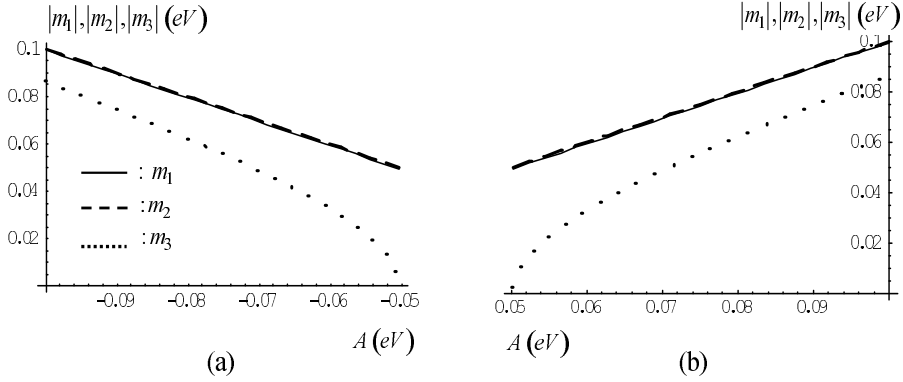}
\vspace*{-0.1cm} \caption[$|m_{1,2, 3}|$ as functions of $A$ in
 the inverted hierarchy with a) $A \in(-0.1, -0.0503) \, \mathrm{eV}$ and 
b) $A\in(0.0503, 0.1) \, \mathrm{eV}$.]{$|m_{1,2, 3}|$ as functions of $A$ in
 the inverted hierarchy with a) $A \in(-0.1, -0.0503) \, \mathrm{eV}$ and 
b) $A\in(0.0503, 0.1) \, \mathrm{eV}$.}\label{m123I}
\ec
\vspace*{-0.4cm}
\end{figure}

\begin{figure}[h]
\begin{center}
\includegraphics[width=12.0cm, height=5.0cm]{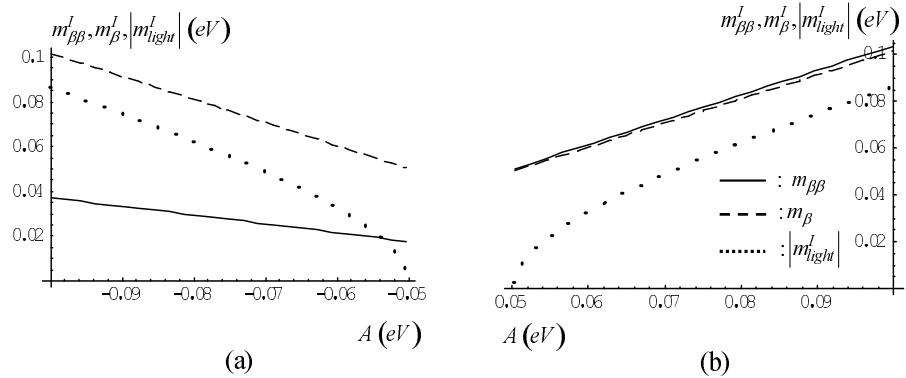}
\caption[$m^I_{\beta}$, $m^I_{\beta\beta}$ and $|m^I_{light}|$ as functions of $A$ in the normal hierarchy with a) $A \in(-0.1, -0.0503) \, \mathrm{eV}$ and 
b) $A\in(0.0503, 0.1) \, \mathrm{eV}$.]{$m^I_{\beta}$, $m^I_{\beta\beta}$ and $|m^I_{light}|$ as functions of $A$ in the normal hierarchy with a) $A \in(-0.1, -0.0503) \, \mathrm{eV}$ and 
b) $A\in(0.0503, 0.1) \, \mathrm{eV}$.}\label{mbbIcase1}
\end{center}
\end{figure}

By taking the two experimental data on squared mass differences of neutrinos for the inverted hierarchy  
given in Ref. \cite{Gonzalez2014}, $\Delta m^2_{21}=7.50\times 10^{-5}\, \mathrm{eV^2}$ and $\Delta m^2_{31}=-2.449\times 10^{-3}\, \mathrm{eV^2}$, we obtain the relations\footnote{We only consider here one solution with $\delta=\frac{\pi}{2}$.} between $m_{1,3}$ and $m_2=A$ as shown in Fig. \ref{m123I}. 
\\
In the inverted hierarchy\footnote{In the inverted spectrum, $m_3 \sim m_{2} << m_{1}$ hence $m_3$ can be
 considered as the lightest neutrino mass.}, $m_{3} \equiv m^I_{light}$ is the lightest neutrino mass, and the effective neutrino mass from tritium beta decay and the neutrino mass obtained from neutrinoless double-beta decays are plotted in Fig. \ref{mbbIcase1}. 

With $A =5.1\times 10^{-2}\, \mathrm{eV}$, we get explicit values of the model
parameters as follows:
 \bea
m_1&\simeq&5.026 \times10^{-2}\, \mathrm{eV},\hs m_2=5.1\times 10^{-2}\,
 \mathrm{eV},\hs m_3\simeq 8.775 \times 10^{-3} \, \mathrm{eV},\\
m_{\beta\beta}^I&\simeq& 5.1786  \times 10^{-2}\, \mathrm{eV},\hs m_{\beta}^I\simeq
5.1063\times10^{-2}\, \mathrm{eV},\hs \sum^I \simeq 0.11003 \, \mathrm{eV}, \label{meeI}\eea
and
\bea
B_{1,2}&=&(2.95171\mp0.760222i)\times 10^{-2} \,\mathrm{eV},\,\,\, C\simeq 2.20914\times 10^{-2} \,\mathrm{eV}.\label{B12Ci}\eea
Now, combining (\ref{ABCM}) and (\ref{B12Ci}) yields \footnote{This system of equations
 has two solutions, however, these solutions differ only by the sign of $m_D$ which has
 no effect in the neutrino oscillation experiments.}:
\bea
M' &=&-0.979204 M, \,\, m_D = 0.139816i \sqrt{M}, \crn
M_1 &=& -0.1138 M,\,\, M_2 = -0.502898 M,\label{M12MpI}\eea
and
 \bea
 x&=&0.139816\sqrt{M}/v_{\varphi}, \,\,
 y=-0.979204 M/v_\chi,\,\,
 z=-0.1138 M/v_{\zeta_1}, \\
v_{\zeta_2}&=&4.41914 v_{\zeta_1}. \label{xyzv}
 \eea
Eq. (\ref{xyzv}) shows that $v_{\zeta_1}$ and $v_{\zeta_2}$ are different  from each other but in the same order of 
magnitude.

 \section{\label{quark} Quark mass}
The quarks content of the model under $[\mathrm{SU}(2)_L,
\mathrm{U}(1)_Y, \mathrm{U}(1)_X,\underline{S}_4]$ symmetries, respectively, given in Tab. \ref{Quarkcon}, where $i=1,2,3$ is a family index of three
lepton families, which are in order defined as the components of
the $\underline{3}$ representations under $S_4$.

\begin{table}[h]
\tbl{\label{Quarkcon} The quark content of the model.}
{\begin{tabular}{@{}ccccccccc@{}} \toprule
   Fields & $Q_{i L}$ &$u_{1R}$&\,\,$u_{2,3R}$\,\,&\,\,$d_{1R}$\,\,&\,\,$d_{2,3R}$ \\
\noalign{\smallskip}\hline\noalign{\smallskip}
$\mathrm{SU}(2)_L$  & $2$ &$1$&$1$&   $1$  &$1$      \\
$\mathrm{U}(1)_Y$  & $1/3$ &$4/3$&$4/3$&$-2/3$  & $-2/3$    \\ 
$\mathrm{U}(1)_X$ & $0$  &$0$&$0$&   $0$  & $0$    \\
$\underline{S}_4$&  $\underline{3}$  &$\underline{1}$& $\underline{2}$&$\underline{1}$&$\underline{2}$  \\ \botrule
\end{tabular} }
\end{table}

The Yukawa interactions are \footnote{Here, $\tilde{\phi} = i\sigma_2 \phi^*=\left(%
\begin{array}{cc}
\phi^0_2 \\
-\phi^-_1 \\
\end{array}%
\right) \sim [2,-1,0, \underline{3}]$, and $\tilde{\phi'} \sim [2,-1,0, \underline{3}']$.}:
\bea -\mathcal{L}_q &=&  h^u_1 (\bar{Q}_{i L}\tilde{\phi})_{\underline{1}}u_{1R}
+ h^u (\bar{Q}_{i L}\tilde{\phi})_{\underline{2}}u_R
+h'^u (\bar{Q}_{i L}\tilde{\phi'})_{\underline{2}}u_R\crn
&+& h^d_1 (\bar{Q}_{i L}\phi)_{\underline{1}}d_{1R}
+ h^d (\bar{Q}_{i L}\phi)_{\underline{2}}d_R
+ h'^d (\bar{Q}_{i L}\phi')_{\underline{2}}d_R+H.c.\label{Yquark}\eea
With the VEV alignments of $\phi$ and $\phi'$ as given in Eqs. (\ref{vevphi}) and (\ref{vevphip}), the mass Lagrangian of quarks reads
\bea -\mathcal{L}^{mass}_q &=&  (\bar{u}_{1L},\bar{u}_{2L},\bar{u}_{3L}) M_u (u_{1R}, u_{2R},
u_{3R})^T+(\bar{d}_{1L},\bar{d}_{2L},\bar{d}_{3L}) M_d (d_{1R},
d_{2R}, d_{3R})^T\crn
&+&H.c, \label{Lqmass}\eea
where the mass matrices for up-and down-quarks are, respectively, obtained as follows
\bea M_u &=&
\left(%
\begin{array}{ccc}
  h^u_1 v & h^u v-h'^u v'  & h^u v+h'^u v'  \\
  h^u_1 v &(h^u v-h'^u v') \om  & (h^u v+h'^u v') \om^2   \\
  h^u_1 v & (h^u v-h'^u v') \om^2 & (h^u v+h'^u v') \om\\
\end{array}%
\right),\label{Mu}\\
M_d &=&
\left(%
\begin{array}{ccc}
  h^d_1 v & h^d v-h'^d v'  & h^d v+h'^d v'  \\
  h^d_1 v &(h^d v-h'^d v') \om  & (h^d v+h'^d v') \om^2   \\
  h^d_1 v & (h^d v-h'^d v') \om^2 & (h^d v+h'^d v') \om\\
\end{array}%
\right).\label{Md}\eea
The structure of the up- and
down-quark mass matrices in Eqs. (\ref{Mu}) and (\ref{Md}) are similar to those in Ref. \citen{quarksv} ,  i.e, in the model under consideration there is no CP
violation in the quark sector. The matrices $M_u$ and  $M_d$ in Eqs. (\ref{Mu}), (\ref{Md}) are , respectively, diagonalized as
\bea
U^{u+}_L M_u U^u_R&=&\mathrm{diag} \left(\sqrt{3}h^u_1 v, \,\, \sqrt{3}(h^u v-h'^u v'), \,\, \sqrt{3}(h^u v+h'^u v')\right)\crn
&\equiv&\mathrm{diag} \left(m_u, \,\, m_c, \,\,m_t\right), \label{muct}\\
U^{d+}_L M_d U^d_R&=&\mathrm{diag} \left(\sqrt{3}h^d_1 v, \,\, \sqrt{3}(h^d v-h'^d v'),
\,\, \sqrt{3}(h^d v+h'^d v')\right)\crn
&\equiv&\mathrm{diag} \left(m_d, \,\, m_s, \,\,m_b\right), \label{mdsb}\eea
where $U^{u}_L=U^{d}_L=U_{L}$, with $U_{L}$ given in (\ref{Uclep}), are the unitary matrices, which couple 
the left-handed up- and down-quarks to those in the mass bases, respectively, and $U^u_R=U^d_R =1$. Therefore, in this case, we get the
quark mixing matrix \bea U_\mathrm{CKM}=U^{d\dagger}_L
U^u_L=1.\label{CKM}\eea 
This is the common property for some models based on discrete symmetry groups \cite{dlsvS4, dlnvS3, vlD4, vlS4, vlS3,vlT7, vD4, vT7, vlkS4, vlA4} and can be seen as an important result of the paper since the experimental quark mixing matrix is close to the unit matrix. A small permutations such as a violation of $S_4$ symmetry due to unnormal Yukawa interactions will possibly providing the desirable quark mixing pattern\cite{vlD4q} . A detailed study on this problem is out of the scope of this work and should be skip. 

In similarity to the charged leptons, the masses of pairs ($c, t$) and ($s, b$) quarks are also separated by the $\phi'$ scalar. The up and down quark masses are $m_u= \sqrt{3}h^u_1 v,\,\,
m_c= \sqrt{3}(h^u v-h'^u v'),\,\, m_t =\sqrt{3}(h^u v+h'^u v'),\,\,
m_d=\sqrt{3}h^d_1 v,\,\, m_s=\sqrt{3}(h^d v-h'^d v'),\,\, m_b =\sqrt{3}(h^d v+h'^d v')$.i.e, 
\bea 
\frac{m_u}{m_d}&=&\frac{h^u_1}{h^d_1},\,\, \frac{m_c}{m_s}=\frac{h^u v-h'^u v'}{h^d v-h'^d v'},\,\, \frac{m_t}{m_b}=\frac{h^u v+h'^u v'}{h^d v+h'^d v'}.\eea
The current mass
values for the quarks are given by \cite{PDG2014}: \bea
m_u&=&2.3^{+0.7}_{-0.5}\ \textrm{MeV},\,\, m_c=1.275\pm 0.025\
\textrm{GeV},\,\, m_t=173.21\pm0.51\pm0.71\, \textrm{GeV},\crn
m_d&=&4.8^{+0.5}_{-0.3}\ \textrm{MeV}, \,\, m_s=95\pm5\
\textrm{MeV},\hs\hs\,\,\,\,\, m_b=4.18\pm 0.03\
\textrm{GeV}.\label{vien3}\eea
With the help of Eqs. (\ref{muct}), (\ref{mdsb})  and (\ref{vien3}) we obtain the followings relations:
\bea
h^u&=&\frac{5.03695\times 10^{10}}{v},\hs h^d = \frac{1.23409\times 10^{9}}{v}, \crn
h'^u&=&\frac{4.96334\times 10^{10}}{v'},\hs h'^d = \frac{1.17924\times 10^{9}}{v'}, \crn
h^u_{1}&=&\frac{1.32791\times 10^{6}}{v},\hs h^d_1 = \frac{2.77128\times 10^{6}}{v},\label{hudrelation}
\eea
or
\bea
h^u/h^d&\simeq&40,\hs h'^u/h'^d\simeq 42, \hs h^d_1/h^u_1\simeq 2,\label{hudratio}
\eea
\bea
h^u/h^u_1&\simeq&3.8\times 10^{4},\hs h^d/h^d_1\simeq 4.5\times 10^{2},\label{hudratio1}
\eea
i.e, $h^u_1$ and $h^d_1$ are in the same order but $h^u \,(h'^u)$ is one magnitude order larger than $h^d\, (h'^d)$. On the other hand, 
in the case $|v| \sim|v'|$ we get $h'^u/h^u_{1}\simeq 3.7\times 10^{4}, \,\, h'^d/h^d_{1}\simeq 4.2\times 10^{2}$. 

To get explicit values of the Yukawa couplings in the quark sector, we assume $v' \sim v \sim 100 \, \mathrm{Gev}$ then 
\bea
h^u&=&0.503695,\hs\hs\,\,\,\, h'^u=0.496334, \hs\hs\,\,\,\, h^u_{1}=1.32791\times 10^{-5},\crn
 h^d &=&1.23409 \times 10^{-2}, \,\, h'^d = 1.17924\times 10^{-2}, \,\, h^d_1 =2.77128\times 10^{-5}.\label{quarks_v}
\eea
We note that, the quarks mixing matrix in Eq. (\ref{CKM}) has no predictive power for quarks mixing but their masses are consistent with the recent experimental data.   
 \section{\label{conclus}Conclusions}
We have proposed a neutrino mass model based on $S_4$ flavor symmetry which accommodates lepton mass, mixing with non-zero $\theta_{13}$ and CP violation phase, and the quark mixing matrix is unity at tree level. The realistic neutrino mass and mixing pattern obtained at the tree- level with renormalizable interactions by one $SU(2)_L$ doubplet and two $SU(2)_L$ singlets  in which one being in $\underline{2}$ and the two others in $\underline{3}$ under $S_4$ if both the breakings $S_{4}\rightarrow S_3$ and $S_{4}\rightarrow Z_3$ are taken place in charged lepton sector and the breaking $S_4\rightarrow \mathcal{K}$ taken place in neutrino sector. The model also gives a remarkable prediction of Dirac CP violation $\delta_{CP}=\frac{\pi}{2}$ or $-\frac{\pi}{2}$ in the both normal and inverted spectrum.

\appendix
\section{\label{S4group}$\emph{S}_4$ group and Clebsch-Gordan coefficients}

For convenience, we will refer to some properties of $S_4$ \cite{dlsvS4}. $S_4$ has 24 elements divided into 5
conjugacy classes, with \underline{1}, \underline{1}$'$,
\underline{2}, \underline{3}, and \underline{3}$'$ as its 5
irreducible representations. Any element of $S_4$ can be formed by
multiplication of the generators $S$ and $T$ obeying the relations
$S^4 = T^3 = 1,\ ST^2S = T$. In this paper, we work in the basis where $\underline{3},\underline{3}'$ are
real representations whereas $\underline{2}$ is complex. One
possible choice of generators is given as follows \bea
\underline{1}&:& S=1,\hs T=1 \crn \underline{1}'
&:& S=-1,\hs T=1 \crn \underline{2} &:& S=\left(%
\begin{array}{cc}
  0 & 1 \\
  1 & 0 \\
\end{array}%
\right),\hs T=\left(%
\begin{array}{cc}
  \om & 0 \\
  0 & \om^2 \\
\end{array}%
\right)\crn \underline{3}&:& S=\left(%
\begin{array}{ccc}
  -1 & 0 & 0 \\
  0 & 0 & -1 \\
  0 & 1 & 0 \\
\end{array}%
\right),\hs T=\left(%
\begin{array}{ccc}
  0 & 0 & 1 \\
  1 & 0 & 0 \\
  0 & 1 & 0 \\
\end{array}%
\right)\crn
\underline{3}'&:& S=-\left(%
\begin{array}{ccc}
  -1 & 0 & 0 \\
  0 & 0 & -1 \\
  0 & 1 & 0 \\
\end{array}%
\right),\hs T=\left(%
\begin{array}{ccc}
  0 & 0 & 1 \\
  1 & 0 & 0 \\
  0 & 1 & 0 \\
\end{array}%
\right)\eea where $\om=e^{2\pi i/3}=-1/2+i\sqrt{3}/2$. 
All the group multiplication
 rules of $S_4$ as given below.
\bea
\underline{1}\otimes\underline{1}&=&\underline{1}(11),\hs
\underline{1}'\otimes \underline{1}'=\underline{1}(11),\hs
\underline{1}\otimes\underline{1}'=\underline{1}'(11),\label{A1}\\
\underline{1}\otimes \underline{2}&=&\underline{2}(11,12),\hs
\underline{1}'\otimes \underline{2}=\underline{2}(11,-12),\\
\underline{1}\otimes \underline{3}&=&\underline{3}(11,12,13),\hs
\underline{1}'\otimes \underline{3}=\underline{3}'(11,12,13),\\
\underline{1}\otimes \underline{3}'&=&\underline{3}'(11,12,13),\hs
\underline{1}'\otimes \underline{3}'=\underline{3}(11,12,13),\\
\underline{2} \otimes \underline{2} &=& \underline{1}(12+21)
\oplus \underline{1}'(12-21) \oplus \underline{2}(22,11),\\
\underline{2}\otimes \underline{3}&=&
\underline{3}\left((1+2)1,\om (1+\om 2)2,\om^2 (1+\om^2 2)
3\right)\crn && \oplus \underline{3}'\left((1-2)1,\om (1-\om
2)2,\om^2 (1-\om^2 2) 3\right) \\ \underline{2}\otimes
\underline{3}'&=& \underline{3}'\left((1+2)1,\om (1+\om 2)2,\om^2
(1+\om^2 2) 3\right)\crn &&\oplus \underline{3}\left((1-2)1,\om
(1-\om 2)2,\om^2 (1-\om^2 2) 3\right),\\ 
\underline{3} \otimes \underline{3} &=& \underline{1}(11+22+33)
\oplus \underline{2}(11+\om^2 22+ \om 33,11+\om 22+ \om^2 33) \crn
&&\oplus \underline{3}_s
(23+32,31+13,12+21)\oplus \underline{3}'_a(23-32,31-13,12-21),\\
 \underline{3}' \otimes
\underline{3}' &=&\underline{1}(11+22+33) \oplus
\underline{2}(11+\om^2 22+ \om 33,11+\om 22+ \om^2 33) \crn
&&\oplus \underline{3}_s
(23+32,31+13,12+21)\oplus \underline{3}'_a(23-32,31-13,12-21),\\
\underline{3} \otimes \underline{3}' &=&\underline{1}'(11+22+33)
\oplus \underline{2}(11+\om^2 22+ \om 33,-11-\om 22-\om^2 33) \crn
&&\oplus \underline{3}'_s (23+32,31+13,12+21)\oplus
\underline{3}_a(23-32,31-13,12-21),\label{A11}\eea where the subscripts $s$
and $a$ respectively refer to their symmetric and antisymmetric
product combinations as explicitly pointed out. In the Eqs. (\ref{A1}) to (\ref{A11}) we have used the notation $\underline{3}(1,2,3)$ which means some
$\underline{3}$ multiplet such as $x=(x_1,x_2,x_3)\sim
\underline{3}$ or $y=(y_1,y_2,y_3)\sim \underline{3}$ and so on. Moreover, the
numbered multiplets such as $(...,ij,...)$ mean $(...,x_i
y_j,...)$ where $x_i$ and $y_j$ are the multiplet components of
different representations $x$ and $y$, respectively. 

The rules to conjugate the representations \underline{1},
\underline{1}$'$, \underline{2}, \underline{3}, and
\underline{3}$'$ are given by \bea
\underline{2}^*(1^*,2^*)&=&\underline{2}(2^*,1^*),\hs
\underline{1}^*(1^*)=\underline{1}(1^*),\hs
\underline{1}'^*(1^*)=\underline{1}'(1^*),\\
\underline{3}^*(1^*,2^*,3^*)&=&\underline{3}(1^*,2^*,3^*),\hs
\underline{3}'^*(1^*,2^*,3^*)=\underline{3}'(1^*,2^*,3^*),\eea
where, for example, $\underline{2}^*(1^*,2^*)$ denotes some
$\underline{2}^*$ multiplet of the form $(x^*_1,x^*_2)\sim
\underline{2}^*$.

\section{\label{S4breaking1} The breakings of $S_4$ by triplet $3$}
For triplets $\underline{3}$ we have the followings alignments:
\begin{itemize}
\item[(1)] The first alignment: $\langle \phi_1\rangle\neq\langle \phi_2\rangle\neq\langle \phi_3\rangle$ then $S_{4}$ is  broken into $\{1\}\equiv\{\mathrm{identity}\}$, i.e. $S_{4}$ is completely broken.
\item[(2)] The second alignment: $0\neq\langle \phi_1\rangle\neq\langle \phi_2\rangle=\langle\phi_3\rangle\neq0$ or $0\neq\langle \phi_1\rangle=\langle \phi_3\rangle\neq\langle \phi_2\rangle\neq0$ or $0\neq\langle \phi_1\rangle=\langle \phi_2\rangle\neq\langle \phi_3\rangle\neq0$ then $S_{4}$ is  broken into $Z_2$ which consisting of the elements \{$1, TSTS^2$\} or \{$1, TSS^2$\} or \{$1, S^2TS$\}, respectively.
\item[(3)] The third alignment: $\langle \phi_1\rangle=\langle
\phi_2\rangle =\langle \phi_3\rangle \neq 0$ then $S_{4}$ is  broken into $S_3$ which consisting of the elements \{$1, T, T^2, TSTS^2, STS^2, S^2TS$\}.
\item[(4)] The fourth alignment: $0=\langle \phi_2\rangle\neq\langle \phi_1\rangle=\langle \phi_3\rangle \neq 0$ or $0=\langle \phi_1\rangle\neq\langle \phi_2\rangle=\langle \phi_3\rangle \neq 0$ or $0=\langle \phi_3\rangle\neq\langle \phi_1\rangle=\langle \phi_2\rangle \neq 0$ then $S_{4}$ is  broken into $Z_2$ which consisting of the elements \{$1, TSTS^2$\} or \{$1, TSS^2$\} or \{$1, S^2TS$\}, respectively.
\item[(5)] The fifth alignment: $0=\langle \phi_2\rangle\neq\langle \phi_1\rangle\neq\langle \phi_3\rangle \neq 0$ or $0=\langle \phi_1\rangle\neq\langle \phi_2\rangle\neq\langle \phi_3\rangle \neq 0$ or $0\neq \langle \phi_1\rangle\neq\langle \phi_2\rangle\neq\langle \phi_3\rangle=0$ then $S_{4}$ is completely broken.
\item[(6)] The sixth alignment: $0\neq \langle \phi_1\rangle\neq \langle\phi_2\rangle=\langle \phi_3\rangle=0$ or $0\neq \langle \phi_2\rangle\neq \langle\phi_3\rangle=\langle \phi_1\rangle=0$
or $0\neq \langle \phi_3\rangle\neq \langle\phi_1\rangle=\langle \phi_1\rangle=0$
 then $S_{4}$ is  broken into  Klein four group $\mathcal{K}$ which consisting of the elements \{$1, S^2,TSTS^2,TST$\} or \{$1, TS^2T^2, STS^2, T^2S$\} or \{$1, T^2S^2T, ST^2, S^2TS$\}, respectively.
\end{itemize}

\section{\label{S4breaking2} The breakings of $S_4$ by triplet $3'$}

For triplets $\underline{3}'$ we have the followings alignments:
\begin{itemize}
\item[(1)] The first alignment: $\langle \phi'_1\rangle\neq\langle \phi'_2\rangle\neq\langle \phi'_3\rangle$ then $S_{4}$ is  broken into $\{1\}\equiv\{\mathrm{identity}\}$, i.e. $S_{4}$ is completely broken.
\item[(2)] The second alignment: $0\neq\langle \phi'_1\rangle\neq\langle \phi'_2\rangle=\langle\phi'_3\rangle\neq0$ or $0\neq\langle \phi'_1\rangle=\langle \phi'_3\rangle\neq\langle \phi'_2\rangle\neq0$ or $0\neq\langle \phi'_1\rangle=\langle \phi'_2\rangle\neq\langle \phi'_3\rangle\neq0$ then $S_{4}$ is  broken into $\{1\}\equiv\{\mathrm{identity}\}$, i.e. $S_{4}$ is completely broken.
\item[(3)] The third alignment: $\langle \phi'_1\rangle=\langle
\phi'_2\rangle =\langle \phi'_3\rangle \neq 0$ then $S_{4}$ is broken into $Z_3$ that consists of the elements \{$1, T, T^2$\}.
\item[(4)] The fourth alignment: $0=\langle \phi'_2\rangle\neq\langle \phi'_1\rangle=\langle \phi'_3\rangle \neq 0$ or $0=\langle \phi'_1\rangle\neq\langle \phi'_2\rangle=\langle \phi'_3\rangle \neq 0$ or $0=\langle \phi'_3\rangle\neq\langle \phi'_1\rangle=\langle \phi'_2\rangle \neq 0$ then $S_{4}$ is  broken into $Z_2$ which consisting of the elements \{$1, T^2S$\} or \{$1, TST$\} or \{$1, ST^2$\}, respectively.
\item[(5)] The fifth alignment: $0=\langle \phi'_2\rangle\neq\langle \phi'_1\rangle\neq\langle \phi'_3\rangle \neq 0$ or $0=\langle \phi'_1\rangle\neq\langle \phi'_2\rangle\neq\langle \phi'_3\rangle \neq 0$ or $0\neq \langle \phi'_1\rangle\neq\langle \phi'_2\rangle\neq\langle \phi'_3\rangle=0$ then $S_{4}$ is completely broken.
\item[(6)] The sixth alignment: $0\neq \langle \phi'_1\rangle\neq \langle\phi'_2\rangle=\langle \phi'_3\rangle=0$ or $0\neq \langle \phi'_2\rangle\neq \langle\phi'_3\rangle=\langle \phi'_1\rangle=0$
or $0\neq \langle \phi'_3\rangle\neq \langle\phi'_1\rangle=\langle \phi'_1\rangle=0$
 then $S_{4}$ is  broken into a four-element subgroup generated by a four-cycle, which consisting of the elements \{$1, S, S^2,S^3$\} or \{$1, TST^2, ST, TS^2T^2$\} or \{$1, TS, T^2ST, T^2S^2T$\}, respectively.
\end{itemize}

\section{\label{S4breaking3} The breakings of $S_4$ by doublet $2$}
\begin{itemize}
\item[(1)] 
 The first alignment:  $\langle \zeta_1\rangle=\langle \zeta_2\rangle$ then $S_4$ is
broken into an eight-element subgroup, which is isomorphic to
$D_4$. 
\item[(2)]  
The second alignment: $\langle \zeta_1\rangle\neq 0=\langle
\zeta_2\rangle$ or $\langle \zeta_1\rangle=0\neq \langle \zeta_2\rangle$ then
$S_4$ is broken into $A_4$ consisting of the identity and the even
permutations of four objects. 

\item[(3)]  
The third alignment: $\langle
\zeta_1\rangle\neq \langle \zeta_2\rangle \neq 0$ then $S_4$ is broken
into a four - element subgroup consisting of the identity and
three double transitions, which is isomorphic to Klein four group $\mathcal{K}$. 
\end{itemize}

\section{\label{Npi2} The solution with $\delta=\frac{\pi}{2}$ in the normal case}
By substituting $B_1$ from
(\ref{B1B2v1}) into (\ref{m123}) and taking the two
experimental data on squared mass differences of neutrinos
given in Ref. \cite{Gonzalez2014}, $\Delta m^2_{21}=7.50\times 10^{-5}\, \mathrm{eV^2}$ and $\Delta m^2_{31}=2.457\times 10^{-3}\, \mathrm{eV^2}$, we get a solution (in [eV]) as follows:
 \bea
C&=&0.5\sqrt{\al-2\sqrt{\beta}},\crn
B_2&=&-0.5\sqrt{4A^2-0.0003}+(1.37741\times 10^{-7}  + 0.34412i) C\crn
&-&0.5\sqrt{(3.52631+3.792\times 10^{-7}i) C^2} , \label{B2CN1}
\eea 
where 
\bea
\al&=&(0.0026169 -2.81407\times 10^{-10}i) +(2.26866 -2.43959\times 10^{-7}i) A^2, \\
\beta&=&-2.2987\times 10^{-7} + 4.94378\times 10^{-14} i +(0.00296843 -6.38415\times 10^{-10}i) A^2\crn
&+&(1.2867 -2.7673\times 10^{-7}i) A^4.  \label{alphabetaN1}
\eea

\end{document}